\documentclass[10pt,twocolumn,letterpaper]{article}

\usepackage[pagenumbers]{cvpr} 
\newcommand{\RNum}[1]{\uppercase\expandafter{\romannumeral #1\relax}}
\usepackage{pifont}
\newcommand{\cmark}{\ding{51}}
\newcommand{\xmark}{\ding{55}}

\definecolor{cvprblue}{rgb}{0.21,0.49,0.74}
\usepackage[pagebackref,breaklinks,colorlinks,allcolors=cvprblue]{hyperref}

\def\paperID{***} 
\def\confName{CVPR}
\def\confYear{2025}

\title{Learning Volumetric Neural Deformable Models to Recover 3D Regional Heart Wall Motion from Multi-Planar Tagged MRI}

\author{
Meng Ye$^{1}$, Bingyu Xin$^{1}$, Bangwei Guo$^{1}$, 
Leon Axel$^{2}$, Dimitris Metaxas$^{1}$\\
$_{}^{1}\textrm{}$Rutgers University, $_{}^{2}\textrm{}$New York University School of Medicine\\
{\tt\small \{my389, bx64, bg654, dnm\}@cs.rutgers.edu }
}

\begin{document}
\maketitle
\begin{abstract}
Multi-planar tagged MRI is the gold standard for regional heart wall motion evaluation. 
However, accurate recovery of the 3D true heart wall motion from a set of 2D apparent motion cues is challenging, due to incomplete sampling of the true motion and difficulty in information fusion from apparent motion cues observed on multiple imaging planes. 
To solve these challenges, we introduce a novel class of volumetric neural deformable models ($\upsilon$NDMs).
Our $\upsilon$NDMs represent heart wall geometry and motion through a set of low-dimensional global deformation parameter functions and a diffeomorphic point flow regularized local deformation field.
To learn such global and local deformation for 2D apparent motion mapping to 3D true motion, we design a hybrid point transformer, which incorporates both point cross-attention and self-attention mechanisms.
While use of point cross-attention can learn to fuse 2D apparent motion cues into material point true motion hints, point self-attention hierarchically organised as an encoder-decoder structure can further learn to refine these hints and map them into 3D true motion.
We have performed experiments on a large cohort of synthetic 3D regional heart wall motion dataset. The results demonstrated the high accuracy of our method for the recovery of dense 3D true motion from sparse 2D apparent motion cues. 
Project page is at \url{https://github.com/DeepTag/VolumetricNeuralDeformableModels}.
\end{abstract}    
\section{Introduction}
\label{sec:intro}
\begin{figure}[t]
\begin{center}
\includegraphics[width=1.0\linewidth]{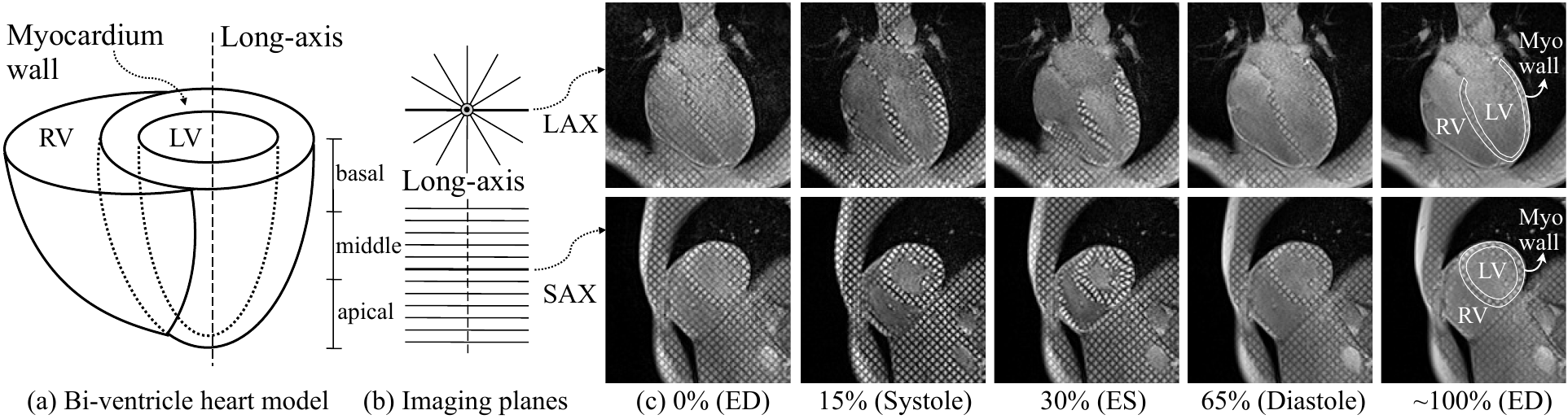}
\end{center}
   \caption{(a) Bi-ventricle heart model. LV: left ventricle. RV: right ventricle. Along the LV long-axis, the heart wall is divided into three parts: basal, middle and apical regions. 
   (b) Long-axis (LAX) (top) and short-axis (SAX) (bottom) imaging planes. 
   (c) Multi-planar tagged MRI sequences of the heart. The number under an image indicates percentage of a cardiac cycle. ED: end diastole. ES: end systole. Myo: myocardium.}
\label{fig1}
\end{figure}

Myocardial tagging with cine magnetic resonance imaging (MRI) provides the gold standard for regional heart wall motion and strain evaluation~\cite{amzulescu2019myocardial, smiseth2024myocardial}. It can also be  used to estimate the twisting motion of myocardium fibers~\cite{lorenz2000function, gotte2006myocardial}.
The spatial modulation of magnetization (SPAMM) technique~\cite{axel1989mr} enables the generation of MRI-visible material landmarks in the wall of left ventricle (LV). As shown in Fig.~\ref{fig1}, the magnetically tagged planes appear as darker tag line patterns embedded in relatively brighter myocardium (Myo) wall. Intersections of two sets of initially orthogonal tagging planes result in traceable material points, which move with the Myo wall deformation during a cardiac cycle. 
Limited by the inherently slow imaging speed of MRI, for the heart, it is more common to perform multi-planar 2D cine imaging than direct 3D$+t$ imaging~\cite{ryf2002myocardial, rutz2008accelerated, ibrahim2011myocardial}. 
To recover the 3D \textbf{true motion} of the Myo wall, we usually shift the imaging plane from the apex to the base of the LV with equal longitudinal intervals to sample a stack of short-axis (SAX) tagging image sequences, and rotate the imaging plane around the long-axis of the LV with equal radial angle intervals to acquire another set of long-axis (LAX) sequences. The SAX and LAX imaging planes are mutually orthogonal. Therefore, the above two sets of tagged MRI data store the in-plane and through-plane \textbf{apparent motion} cues, respectively, which can be jointly utilized to estimate the true motion of the heart wall in the 3D space~\cite{park1996analysis}.  

The 3D heart wall motion recovery from multi-planar 2D tagged MRI is a complex task~\cite{frangi2001three}. It involves the following steps: (1) 2D in-plane motion tracking~\cite{ye2021deeptag, ye2023sequencemorph}; (2) Myocardium wall segmentation~\cite{gao2021utnet, ye2024unsupervised}; (3) Respiratory motion correction~\cite{chang2021unsupervised}; (4) 3D heart wall geometry reconstruction~\cite{ye2023neural}; and (5) 3D motion recovery~\cite{park1996analysis}. Recent advancements in deep learning-based medical image analysis have greatly advanced all of these areas except the last one. Previous efforts on the 3D heart wall motion recovery from tagged MRI can be summarized into the following categories: (1) Finite element model (FEM) approaches~\cite{young1992three, young1994three};  (2) Spline interpolation-based methods~\cite{huang1999spatio}; (3) Physics-based deformable model-inspired approaches~\cite{park1996deformable, park1996analysis, wang2008meshless, yu2014deformable, wang2015meshless}; (4) Others~\cite{o1995three, pan2005fast}. All of these methods rely on iterative optimization, which is time-consuming and prone to a suboptimal convergence result. 

Within these methods, only the deformable model-based approach provides detailed 3D heart wall motion~\cite{park1996analysis, wang2008meshless}, especially for twisting, which can help clinicians for intuitive interpretation. 
We thus adapt this kind of approach, and propose volumetric neural deformable models ($\upsilon$NDMs) to represent the heart wall geometry and recover 3D wall motion from  2D apparent motion cues. 
NDMs have been proposed to reconstruct heart wall geometry~\cite{ye2023neural}, which is intuitive and straightforward, since we only need to guarantee the geometry accuracy.
However, directly applying NDMs to the 3D heart wall motion recovery is challenging, as they cannot accurately reconstruct the \textbf{correspondence} of each material point between different time points.
Our $\upsilon$NDMs can be used to generate synthetic heart wall motion data with known 3D motion ground truth. These synthetic data can then be used to train $\upsilon$NDMs to learn the mapping from apparent motion to true motion.
Material point correspondence could be automatically constructed through sequential motion estimation over time. 


We summarize our key contributions as following: 
\begin{itemize}
\item We propose novel learnable $\upsilon$NDMs to recover 3D true motion from 2D apparent motion cues. While we target 3D regional heart wall motion recovery from 2D tagged MRI sequences, our approach has the potential to be generalized to other time-resolved cardiac MRI applications, such as tissue phase mapping~\cite{jung2006detailed, foll2009visualization} and displacement encoding with stimulated echoes~\cite{aletras1999dense, kim2004myocardial}. 
\item We propose a simulation framework to synthesize complex 3D heart wall geometry and motion during a full cardiac cycle. Synthetic data could be used to train $\upsilon$NDMs for 3D heart wall motion estimation and, more importantly, to quantitatively evaluate the 3D motion recovery accuracy with known underlying ground truth. 
\item We propose a novel hybrid point transformer architecture with point cross-attention and self-attention, in order to overcome specific challenges in dense 3D true motion recovery from sparse 2D apparent motion cues. 


\end{itemize}

\begin{figure}[t]
\begin{center}
\includegraphics[width=1.0\linewidth]{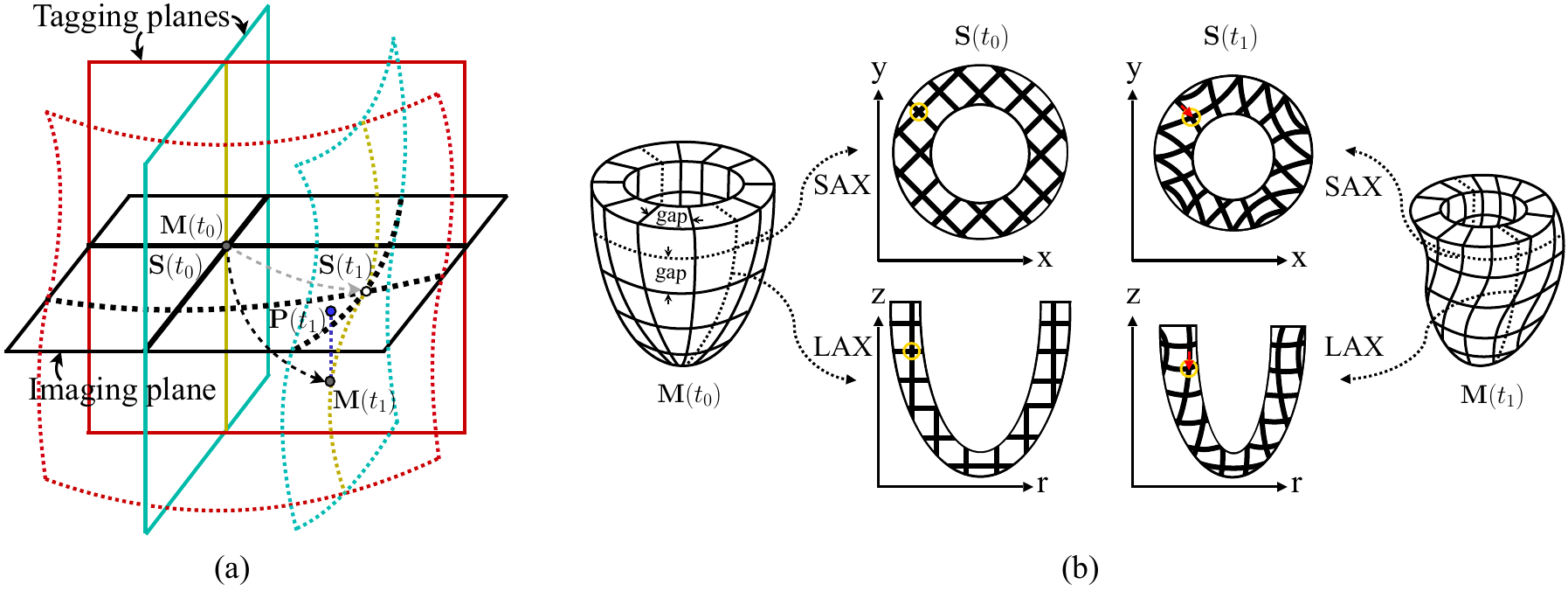}
\end{center}
   \caption{(a) Material points $\mathbf{M}$ and SPAMM datapoints $\mathbf{S}$. 
   (b) In-plane ($x$ and $y$) apparent motion cues (red arrow) provided by two corresponding SAX SPAMM datapoints (yellow) and through-plane ($z$) apparent motion cue (red arrow) provided by two corresponding LAX SPAMM datapoints (yellow).
   The dashed lines in the mesh $\mathbf{M}(t_{1})$ show the SAX and LAX imaging planes identical to those in the mesh $\mathbf{M}(t_{0})$. 
   `r': radius direction.
   }
\label{fig2}
\end{figure}

\section{Background}
\subsection{Material Points and SPAMM Datapoints}
\label{secback}
As shown in Fig.~\ref{fig2} (a), tagging planes created with SPAMM are initially orthogonal to the imaging plane. The intersection line of a tagging plane and the imaging plane appears as a tagging line. At time $t_{0}$, the intersection point of two tagging lines is a \textbf{material point} $\mathbf{M}(t_{0})$. Tagging planes deform with the moving heart wall, so the material point thereby travels to a new location $\mathbf{M}(t_{1})$, which may move out of the \texttt{fixed} imaging plane because of through-plane heart wall motion.  At time $t_{1}$, the new intersection point of the two deformed tagging lines in the \texttt{fixed} imaging plane is the \textbf{SPAMM datapoint} $\mathbf{S}(t_{1})$. Note that, at time $t_{0}$, the SPAMM datapoint coincides with the material point: $\mathbf{S}(t_{0})=\mathbf{M}(t_{0})$. The 3D path between two consecutive material points is the 3D \textbf{true motion} of the heart wall, as shown by the black arrow; the 2D path between two consecutive SPAMM datapoints is the 2D \textbf{apparent motion}, as shown by the gray arrow. 

Because the imaging planes are \texttt{fixed} during a cardiac cycle, we can only rely on the apparent motion cues distributed in multi-planar SPAMM datapoints to recover the 3D true heart wall motion. As shown in Fig.~\ref{fig2} (b), while the SAX SPAMM datapoints provide in-plane ($x$ and $y$) apparent motion cues, the LAX SPAMM datapoints can provide through-plane ($z$) apparent motion cues. 
It is important to note, however, that the SPAMM datapoints in the two orthogonal sets of imaging planes do not correspond to the same material points. On the one hand, this fact inspires us to track the 3D true motion sequentially from the starting time point to the end; with each time we estimate the motion between two consecutive temporal points.
On the other hand, it brings two main challenges for 3D true heart wall motion recovery from 2D apparent motion cues in practice: 
(1) There exist imaging gaps between SAX and LAX tagged MRI slices, respectively. Therefore, the multi-planar SPAMM datapoints are incomplete and sparse samplings of the underlying 3D true heart wall motion. 
(2) Because local 3D heart wall motion is heterogeneous, the projection point $\mathbf{P}(t_{1})$ of $\mathbf{M}(t_{1})$ would not necessarily coincide with $\mathbf{S}(t_{1})$. Therefore, even if we can upsample the sparse apparent motion cues for each material point, it is implausible to directly combine the three $x$, $y$ and $z$ apparent motion components together as the 3D true motion between two consecutive time points. Instead, we need a dedicated apparent motion fusion method to recover the 3D true motion. 

\subsection{Recovering 3D True Heart Wall Motion Using Apparent Motion Cues}
Ever since the introduction of myocardial tagged MRI, numerous efforts have been devoted to the 3D true motion recovery~\cite{frangi2001three, axel2005tagged, wang2011cardiac}. We briefly review them as follows: (1) The finite element model (FEM) approach~\cite{young1992three, young1994three} represents the heart wall by a FEM with first-order continuity~\cite{nielsen1991mathematical}. The continuous FEM of the heart wall is deformed to fit the displacement field of the SPAMM datapoints. A smoothness constraint is applied to the deformed FEM which regularizes motion interpolation between SPAMM datapoints and imaging planes.  However, large motion recovering errors could occur in the imaging gaps, because of simplified interpolation functions, \eg, Hermite basis functions~\cite{hunter1988analysis}.  
(2) Spline interpolation-based methods~\cite{moulton1996spline, huang1999spatio, ozturk2000four, deng2004three, luo2005lv, segars2009improved, wu2013cardiac, henn2015dilated} use 3D B-solids to model the heart wall, which are deformed to fit the tagging planes or SPAMM datapoints across time. The optimization procedure of 3D motion fitting can be formulated with additional smoothness or incompressibility constraints. The motion vectors estimated in imaging gaps are the 3D B-spline interpolation results of tracked motion vectors at the control points of the B-solid model. 
(3) In the deformable model-~\cite{terzopoulos1991dynamic} inspired approach~\cite{park1996deformable, park1996analysis, wang2008meshless, chen2009automated, yu2014deformable, wang2015meshless}, the heart wall is represented by volumetric deformable models. 
At first, volumetric deformable models were represented by finite element models~\cite{park1996analysis, park1996deformable} and later by meshless deformable models~\cite{wang2008meshless, wang2015meshless}, to avoid remeshing procedures when computing large deformations.
The 3D heart wall motion is described by a set of locally varied deformation parameter functions~\cite{park1996deformable}, which are piecewise linear along the LV long-axis and thus cannot describe further details of heterogeneous local wall motion.  
(4) Others, including iterative
point-tracking technique~\cite{moore1992calculation, zhong2008characterization}, the discrete model-free algorithm~\cite{denney1995reconstruction, denney1997model}, displacement field fitting~\cite{o1995three, moore2000three}, 3D nonrigid image registration~\cite{chandrashekara2004analysis, oubel2012cardiac}, and 3D harmonic phase (HARP) tracking~\cite{pan2005fast}.


\section{Method}
\subsection{Principles of Volumetric Neural Deformable Models ($\upsilon$NDMs)}
\label{sec31preliminary}

\begin{figure}[t]
\begin{center}
\includegraphics[width=1.0\linewidth]{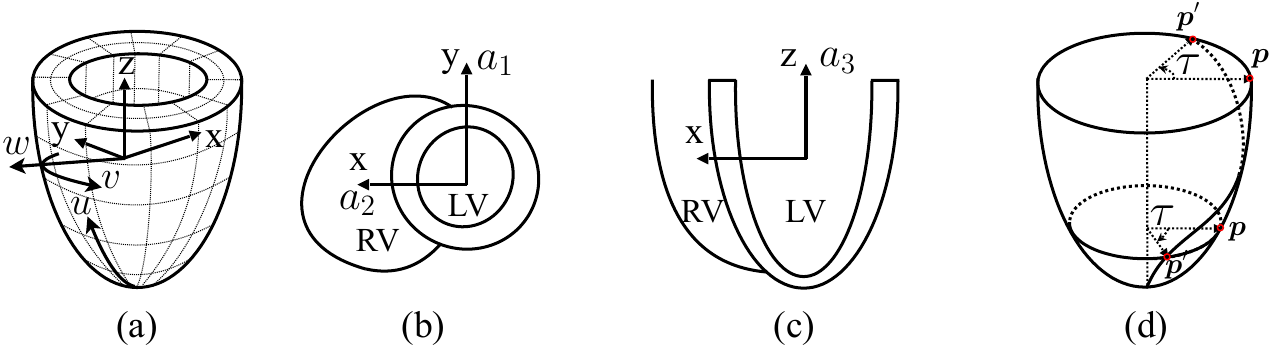}
\end{center}
   \caption{(a) Coordinate system definition of a volumetric deformable model. 
   (b) Short-axis view.
   (c) Long-axis view.
   (d) Twisting deformation $\tau$ that rotates an in-plane point from $\boldsymbol{p}$ to $\boldsymbol{p}'$. 
   }
\label{fig3}
\end{figure}

\begin{figure*}[t]
\begin{center}
\includegraphics[width=0.98\linewidth]{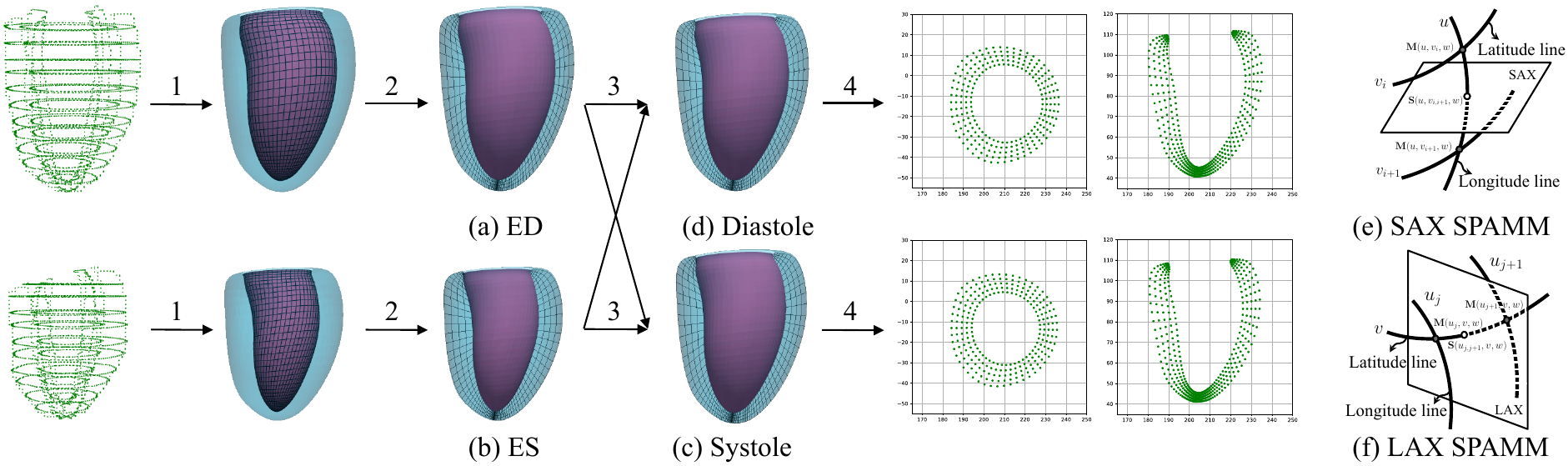}
\end{center}
   \caption{Steps for heart wall geometry and motion simulation, and SPAMM datapoints computation. We first fit the inner and outer wall at (a) ED and (b) ES from a sparse point cloud, using a two-layer $\upsilon$NDM in step $1$, and then interpolate middle wall layers in step $2$. We synthesize missing heart wall models at other temporal points (\eg, (c) systole and (d) diastole) by interpolating the models at ED and ES phases in step $3$. In step $4$, we compute the SPAMM datapoints in (e) SAX and (f) LAX views using a mesh-plane clipping algorithm. 
   }
\label{fig4}
\end{figure*}

We represent the LV heart wall geometry and motion  with a volumetric neural deformable model ($\upsilon$NDM), which extends NDM~\cite{ye2023neural} from two surfaces to multiple surfaces and augments the deformation parameters with twisting. 
We first define an inertial reference frame $\boldsymbol{\Phi }$ in 3D space. 
Material points $\mathbf{M}$ in the model, relative to $\boldsymbol{\Phi }$, are expressed as vector-valued, time-varying functions of material coordinates $\boldsymbol{m}=\left ( u, v, w \right )$: $\mathbf{M}(\boldsymbol{m}, t)=\left ( x(\boldsymbol{m}, t), y(\boldsymbol{m}, t), z(\boldsymbol{m}, t) \right )^{\top}$, where $t$ represents time and $\top$ is the transpose operator.
We further introduce a model-centered reference frame $\boldsymbol{\phi }$, whose origin is $\boldsymbol{c}$, and whose orientation is described by a rotation matrix $\boldsymbol{R}$ represented as a quaternion with unit magnitude. 
As shown in Fig.~\ref{fig3} (a)(b)(c), at ED phase, short- and long-axis views coincide with $xy$- and $yz$-planes in the model frame $\boldsymbol{\phi }$, respectively. Material coordinate $u$ runs from the apex to the base of the LV, $v$ starts and ends at the point where the mid septum wall is located, and $w$ is used to define intermediate wall layers between inner and outer walls.

The heart wall geometry is defined through a shape primitive $\boldsymbol{s}$ in the model frame $\boldsymbol{\phi }$, and the points $\mathbf{M}$ can be represented as
\begin{equation}    
\mathbf{M}=\boldsymbol{c}+\boldsymbol{R}(\boldsymbol{s}+\boldsymbol{d}),
\label{eq_qcsd}
\end{equation}
where $\boldsymbol{d}$ describes local deformations. 
The shape primitive $\boldsymbol{s}$ is defined by an ellipsoid with \textit{parameter functions} ~\cite{park1996deformable} of $u, w$, accounting for a certain amount of local variation:
\begin{equation}
\begin{aligned}
\boldsymbol{e}=
a_{0}(w)\begin{pmatrix}
a_{1}(u, w)\, cos\, u\, cos\, v\\ 
a_{2}(u, w)\, cos\, u\, sin\, v\\ 
a_{3}(u, w)\, sin\, u\\ 
\end{pmatrix}.
\end{aligned}
\label{eq_pf}
\end{equation}
where $-\pi /2\leq u\leq \pi /6$, $-\pi \leq v < \pi$,  $w=0, 1, ..., N_{w}-1$, $a_{0}> 0$, $0\leq a_{1},a_{2},a_{3}\leq 1$. In Eq.~(\ref{eq_pf}), $a_{0}$ is the scale parameter, and $a_{1}$, $a_{2}$ and $a_{3}$ are the aspect ratios along $x$-, $y$- and $z$-axis, respectively. 

We further define and apply the twisting deformation \textit{parameter functions} $\tau(u, w)$ along the $z$-axis, as shown in Fig.~\ref{fig3} (d), and the axis offsets $(e_{xo}(u, w)$, $e_{yo}(u, w))$ to the above ellipsoid primitive $\boldsymbol{e}=\left (e_{x}, e_{y}, e_{z}\right )^{\top}$. The resulting shape primitive $\boldsymbol{s}$ can be expressed as:
\begin{equation}
\boldsymbol{s}=\begin{pmatrix}
e_{x}cos(\tau(u,w))-e_{y}sin(\tau(u,w))+e_{xo}(u, w)\\ 
e_{x}sin(\tau(u,w))+e_{y}cos(\tau(u,w))+e_{yo}(u, w)\\ 
e_{z}
\end{pmatrix}.
\label{eq_tw}
\end{equation}
Note that the \textit{parameter functions} are defined with local $(u, w)$, and ``globally" govern a local set of points on $\left\{(u, v_{i},w) \right\}_{i=0}^{N_{v}-1}$. Hence, we call them global parameters.

While previous deformable model-based methods~\cite{park1996deformable, park1996analysis} ignore local deformations $\boldsymbol{d}$, we follow~\cite{ye2023neural} to define those deformations as diffeomorphic point flows~\cite{younes2010shapes, brezis2011functional}, aiming at preserving the topology of $\upsilon$NDMs. 
For $N_{m}$ material points $\mathbf{M}$ with initial positions at $\mathbf{M}^{(0)}$, we define a smooth \textit{velocity field} $\boldsymbol{v}:\mathcal{V} \subset \mathbb{R}^{N_{m}\times 3}\times [0, 1] \mapsto \mathcal{V} \subset \mathbb{R}^{N_{m}\times 3}$, which describes the trajectory of a spatial mapping  $\mathcal{D}(\mathbf{M}, h):\mathcal{V} \subset \mathbb{R}^{N_{m}\times 3}\times [0, 1] \mapsto \mathcal{V} \subset \mathbb{R}^{N_{m}\times 3}$ parameterized by time $h\in \left [ 0,1 \right ]$ through the following flow equation~\cite{dupuis1998variational}: 
\begin{equation}
\frac{\partial \mathcal{D}(\mathbf{M}, h)}{\partial h}=\boldsymbol{v}(\mathcal{D}(\mathbf{M}, h), h)\; \;\; \; s.t.\; \;\; \; \mathcal{D}(\mathbf{M}, 0)=\mathbf{M}^{(0)},
\label{eq_flow}
\end{equation}
where $\mathbf{M}^{(1)}=\mathcal{D}(\mathbf{M}, 1)$ are final transformed points, and $\boldsymbol{d}=\mathcal{D}(\mathbf{M}, 1)-\mathcal{D}(\mathbf{M}, 0)$. 
By setting the dynamic function as the velocity field $\boldsymbol{v}$, we can solve the ordinary equation (ODE) in Eq.~(\ref{eq_flow}) with a neural ODE (NODE) solver~\cite{chen2018neural}.

\begin{figure*}[t]
\begin{center}
\includegraphics[width=0.91\linewidth]{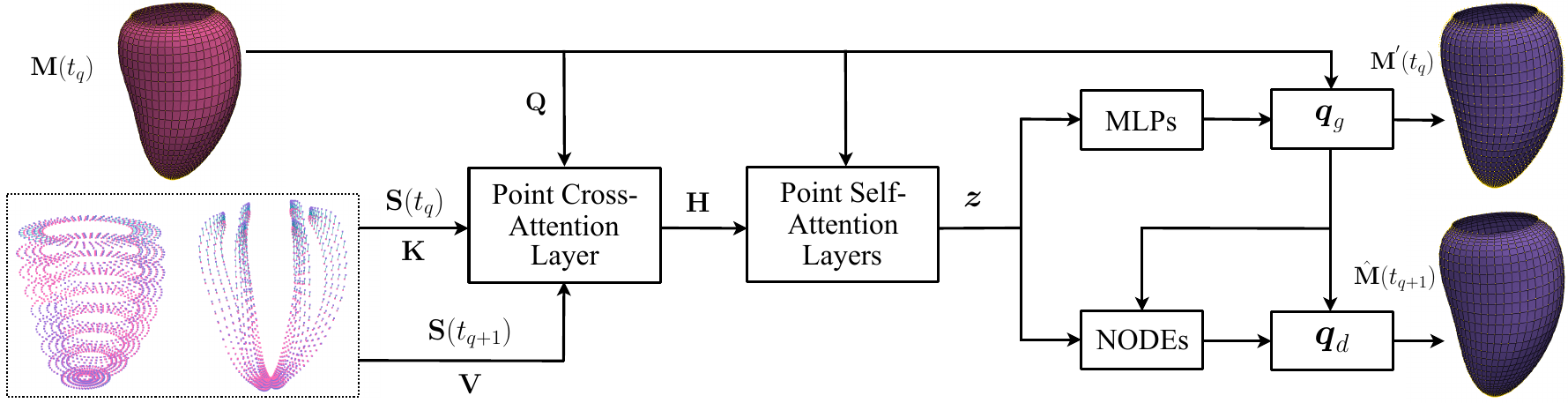}
\end{center}
   \caption{Heart wall 3D motion recovery network. The point cross-attention layer fuses sparse apparent motion cues $\mathbf{A}(t_{q},  t_{q+1}) = \left\{\mathbf{S}(t_{q}), \mathbf{S}(t_{q+1})\right\}$ into true motion hints $\mathbf{H}$. Multiple point self-attention layers transform true motion hints $\mathbf{H}$ to latent motion code $\boldsymbol{z}$, which is further used by the MLPs and conditional neural ordinary diffeomorphic blocks (NODEs) to predict 3D true motion of material points. 
   In the dashed box, pink points show $\mathbf{S}(t_{q})$; purple points show $\mathbf{S}(t_{q+1})$; lines between $\mathbf{S}(t_{q})$ and  $\mathbf{S}(t_{q+1})$ show one-to-one correspondences.
   Yellow points on/near the mesh of the middle myocardium wall layer show ground truth material points. 
   }
\label{fig5}
\end{figure*}

\subsection{Simulation of 3D Wall Geometry and Motion }
\label{simulation}
We use a 3D cardiac cine MRI dataset~\cite{bai2015bi} to simulate left ventricular wall geometry and regional motion over a cardiac cycle. Each subject in the dataset has 3D heart wall segmentation masks on two phases (ED and ES). As shown in Fig.~\ref{fig4}, our simulation process has the following steps: 
\begin{itemize}
\item Step $\boldsymbol{\mathit{1}}$: We train a two-layer $\upsilon$NDM to fit the inner and outer wall geometry for each heart wall instance without activation of the twisting parameters. During inference, for the ED phase, the twisting parameter function is set as zero: $\tau_{ed}(u, w)=0$. For the ES phase, we set $\tau_{es}(u, w)$ according to the normal subject value in~\cite{park1996analysis}. As shown in~\cite{ye2023neural}, using NDM to model heart wall geometry, we can easily reconstruct heart wall meshes and build dense correspondence between different shape instances. 
\item Step $\boldsymbol{\mathit{2}}$: We discretize each $\upsilon$NDM by linearly interpolating between the corresponding material points on inner and outer walls: 
\begin{equation}
\mathbf{M}(u, v, w) =  s_{w}\mathbf{M}(u, v, w_{out}) + (1 - s_{w})\mathbf{M}(u, v, w_{in}),
\end{equation}
where $s_{w}=\frac{w}{N_{w} - 1}$, $w = 0, \dots, N_{w} - 1$, $w_{in} = 0$, $w_{out} = N_{w} - 1$.
This step will produce $N_{w}-2$ intermediate wall layers, which are connected with the inner and outer walls to formulate the complete volumetric mesh.
\item Step $\boldsymbol{\mathit{3}}$: We generate the missing deforming $\upsilon$NDMs at other $T-2$ temporal points in a cardiac cycle by interpolating between all the corresponding material points of the ED and ES phases:
\begin{equation}
\mathbf{M}(t_{q}) =  s(t_{q})\mathbf{M}(t_{es}) + (1-s(t_{q}))\mathbf{M}(t_{ed}),
\label{eqmtq}
\end{equation}
where $s(t_{q})\in[0, 1]$, $s(t_{ed}) = 0$, $s(t_{es}) = 1$, $q = 0, \dots, T-1$, $t_{ed} = t_{0}$, $t_{es}\approx t_{T/3}$. As such, we can synthesize a $\upsilon$NDM sequence of $T$ frames for each subject, which represents the trajectory of every material point in the heart wall in a cardiac cycle. More details of $s(t_{q})$ setting can be found in the Supplementary Material.
\item Step $\boldsymbol{\mathit{4}}$: We compute the SPAMM datapoints in the SAX and LAX views by using a mesh-plane clipping algorithm~\cite{foley1996computer}. The definition of SAX and LAX views can be found in Fig.~\ref{fig1}. The intersection points of the imaging planes with the $\upsilon$NDM quadrilateral mesh are calculated as the SPAMM datapoints, which are 3D points\footnote{In Fig.~\ref{fig4}, we show the 2D projections of SPAMM datapoints on their imaging planes.} as shown in Fig.~\ref{fig6} (a)(b).
As shown in Fig.~\ref{fig4} (e)(f), along a longitude or latitude material line, two closest material points to the imaging plane are determined first; then they are linearly interpolated as the intersection point. 
Material coordinates $(u, v, w)$ of SPAMM datapoints are recorded for correspondence retrieval.
\end{itemize}

\subsection{Recovering 3D Heart Wall Motion with $\upsilon$NDMs}
Our 3D true motion recovery network $\Psi$ is shown in Fig.~\ref{fig5}. In essence, it is a hybrid point transformer which maps sparse apparent motion cues into dense true motion between two consecutive time points. The input of  $\Psi$ includes two parts: 
(1) Material points $\mathbf{M}(t_{q})$ at a certain \textit{query} time point $t_{q}$; (2) Apparent motion cues $\mathbf{A}(t_{q}, t_{q+1})$ between $t_{q}$ and $t_{q+1}$, which are the corresponding SPAMM datapoint pairs: $\mathbf{A}(t_{q},  t_{q+1}) = \left\{\mathbf{S}(t_{q}), \mathbf{S}(t_{q+1})\right\}$.
We adapt the point attention mechanism proposed in Point Transformer~\cite{zhao2021point} for our task. For a query point $\mathbf{Q}_{i}$, we assign points in its $k$-nearest neighbour ($k$-NN) $\mathcal{N}(i)$ as key points $\mathbf{K}$, and corresponding features of key points as values $\mathbf{V}$. The vector attention with the subtraction relation is used to transform $\mathbf{Q}_{i}$,  $\mathbf{K}$ and $\mathbf{V}$ into output point feature $\mathbf{Y}_{i}$:
\begin{equation}
\mathbf{Y}_{i}=\sum_{j\in \mathcal{N}(i)}\rho (\gamma (\varphi (\mathbf{Q}_{i})-\psi (\mathbf{K}_{j})+\delta ))\odot (\alpha (\mathbf{V}_{j})+\delta),
\end{equation}
where $\rho$ is the \textit{softmax} function, $\varphi$, $\psi$, and $\alpha$ are linear transformations, $\gamma$ is a mapping function, and $\delta$ is a position encoding. 
More details of point attention mechanism and implementation can be found in~\cite{zhao2021point}. 

\begin{figure}
\begin{center}
\includegraphics[width=0.92\linewidth]{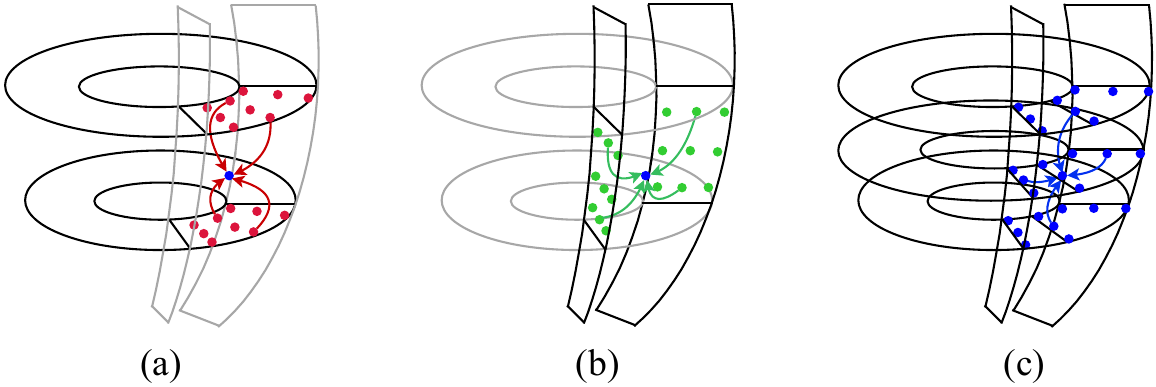}
\end{center}
   \caption{Query points and key points in point cross-attention and self-attention.
   A query point (blue) is always a material point. In (a) (b) cross-attention, key points are the SPAMM datapoints either on SAX views (red) or LAX views (green).  For (c) self-attention, key points are neighbouring material points (blue). Material points and SPAMM datapoints do not necessarily coincide with each other.
   }
\label{fig6}
\end{figure}

We first use a point cross-attention layer to fuse apparent motion cues. As shown in Fig.~\ref{fig6} (a)(b), we set query points as material points $\mathbf{M}(t_{q})\in \mathbb{R}^{N_{m}\times 3}$, key points as SPAMM datapoints $\mathbf{S}(t_{q})\in \mathbb{R}^{N_{s}\times 3}$, valves as corresponding SPAMM datapoints $\mathbf{S}(t_{q+1})\in \mathbb{R}^{N_{s}\times 3}$. 
$\mathbf{S}(t_{q})$ and  $\mathbf{S}(t_{q+1})$ have one-to-one \textbf{correspondences}, as indicated in Fig.~\ref{fig5}.
$N_{m}$ and $N_{s}$ are material point and SPAMM datapoint numbers, respectively.
$N_{m}=N_{u}\times N_{v} \times N_{w}$, where $N_{u}$, $N_{v}$ and $ N_{w}$ are the numbers of material coordinates $u$, $v$, $w$, respectively. 
The SAX and LAX views are processed in parallel.
The output point features of SAX and LAX views are concatenated at the feature channel dimension and fed into an two-layer miltilayer perceptron (MLP), to further mix and fuse the apparent motion cues from different planes. 
Using the point cross-attention with an MLP, we can fuse apparent motion cues from different 2D planes into 3D true motion hints $\mathbf{H}$. 
By querying each material point in $\mathbf{M}(t_{q})$ with true motion hints, the cross-attention layer has the effect of upsampling sparse apparent motion cues as dense as the querying material points. 

We then use multiple point self-attention layers of an encoder-decoder structure, as in Point Transformer~\cite{zhao2021point}, to further refine 3D true motion hints $\mathbf{H}$ and transform them into latent motion code $\boldsymbol{z}\in \mathbb{R}^{N_{m}\times C_{z}}$. As shown in Fig.~\ref{fig6} (c), in point self-attention layers, the query, key and value points are identical.

All the geometry and deformation paremeters in $\upsilon$NDMs could be used to describe the heart wall motion. However, during 3D motion recovery, we only allow $\boldsymbol{q}_{g} = (\boldsymbol{a}_{1}, \boldsymbol{a}_{2}, \boldsymbol{a}_{3}, \boldsymbol{\tau})^{\top}$ and $\boldsymbol{q}_{d}=\boldsymbol{d}$ deformable~\cite{wang2015meshless}. This is plausible because: (1) The heart motion between two consecutive time points is small; (2) The deformation effects of $(\boldsymbol{C}, \boldsymbol{a}_{0})$, $\boldsymbol{R}$, and $(\boldsymbol{e}_{xo}, \boldsymbol{e}_{yo})$, if any, could be counted to $(\boldsymbol{a}_{1}, \boldsymbol{a}_{2}, \boldsymbol{a}_{3})$, $\boldsymbol{\tau}$, and $\boldsymbol{q}_{d}$, respectively. 

With $\boldsymbol{z}$, we first predict the global deformation vectors:
\begin{equation}
\boldsymbol{q}_{g}(u, w) = \text{MLP}(\boldsymbol{z}(u, w)),
\end{equation}
where a certain global deformation vector $\boldsymbol{q}_{g}(u, w) \in \mathbb{R}^{4 \times 1}$ is mapped from the corresponding latent motion code $\boldsymbol{z}(u, w)$, and we deform $\mathbf{M}(t_{q})$ as $\mathbf{M}^{'}(t_{q}) = \mathbf{M}(t_{q}) \circ \boldsymbol{q}_{g}$. 
During this global deformation of local material point sets $\mathbf{M}^{'}(u,.,w; t_{q})$ governed by each  $(u, w)$, we take $\mathbf{M}$ as $\boldsymbol{e}$ in Eq.~(\ref{eq_pf})(\ref{eq_tw}).
Then we use conditional NODEs~\cite{ye2023neural, gupta2020neural} to predict the local deformation field  $\boldsymbol{q}_{d}\in \mathbb{R}^{N_{m}\times 3}$:
\begin{equation}
\frac{\partial \mathcal{D}(\mathbf{M}; \boldsymbol{z}, h)}{\partial h}=\boldsymbol{v}(\mathcal{D}(\mathbf{M}; \boldsymbol{z}, h); \boldsymbol{z}, h),
\label{eq_Cflow}
\end{equation}
where $\mathcal{D}(\mathbf{M}; \boldsymbol{z}, 0)=\mathbf{M}^{'}(t_{q})$
and we get the final deformed material points as $\hat{\mathbf{M}}(t_{q+1})= \mathbf{M}^{'}(t_{q}) \circ \boldsymbol{q}_{d}$. 

In summary, the deformation parameters in $\upsilon$NDM to be recovered are $\boldsymbol{q}_{N}^{\upsilon }=(\boldsymbol{q}_{g}^{\top}, \boldsymbol{q}_{d}^{\top})^{\top}$. The 3D motion recovering network $\Psi$ achieves the following mapping:
\begin{equation}
\Psi(\mathbf{M}(t_{q}), \mathbf{A}(t_{q}, t_{q+1}))) \mapsto (\hat{\mathbf{M}}(t_{q+1}), \boldsymbol{q}_{N}^{\upsilon}(t_{q}, t_{q+1})).
\label{eq_Psimapping}
\end{equation}

\subsection{Model Training}
We train the 3D true motion recovery network $\Psi$ using the sum of $\ell_{2}$ loss and two regularization losses: 
\begin{equation}
\mathcal{L} = \ell_{2}(\hat{\mathbf{M}}(t_{q+1})), \mathbf{M}(t_{q+1}))+\lambda_{d} \mathcal{L}_{d}+\lambda_{s} \mathcal{L}_{s},
\end{equation}
where $\mathbf{M}(t_{q+1})$ are ground truth material points at $t_{q+1}$, $\lambda_{d}$ and $\lambda_{s}$ are the weighting hyper-parameters.
Following~\cite{ye2023neural}, we use $\mathcal{L}_{d}=\left \| \boldsymbol{q}_{d} \right \|_{2}^{2}$ to regularize the amount of local deformations and use $\mathcal{L}_{s}=\left \| \bigtriangledown  \boldsymbol{q}_{d} \right \|_{2}^{2}$ to regularize the smoothness of the local deformation field.

Our task is to recover deformation parameters $\boldsymbol{q}_{N}^{\upsilon }$ in $\upsilon$NDM. Since $\boldsymbol{q}_{N}^{\upsilon }$ contains a large set of deformation \textit{parameter functions}, simultaneously optimization of them will result in a large search space and lead to slow convergence speed.
We use the marginal space learning (MSL) method~\cite{zheng2007fast, raju2022deep, ye2023neural} to train $\Psi$, in which we gradually increase the deformation freedom of $\boldsymbol{q}_{N}^{\upsilon }$ into $\upsilon$NDM for a sub period of training epochs.

As illustrated next, during inference, we take initial material points $\mathbf{M}(t_{0})$ to predict material points $\mathbf{M}(t)$ at all remaining time points in a cardiac cycle. Accumulation error could make the prediction $\hat{\mathbf{M}}(t)$ gradually drift from the ground truth $\mathbf{M}(t)$. To reduce such accumulation error, we propose a two-stage training scheme:
\begin{itemize}
\item Stage \RNum{1}: We randomly sample two consecutive training data sets ($L_{t}=2$) for a subject, and use the MSL method to train $\upsilon$NDMs for $E_{1}$ epochs;
\item Stage \RNum{2}: We increase the consecutive training data time length to $L_{t}=5$ for a subject, and use the sequential prediction method illustrated next to account for accumulation error. This stage lasts for $E_{2}$ epochs, to refine $\upsilon$NDMs and make them robust on possible motion estimation drift.
\end{itemize}

\subsection{Model Inference}
During inference of the 3D true motion in a cardiac cycle for a subject, we are given the input of initial material points at ED phase and apparent motion cues at each subsequent temporal points.
We always start from $t_{ed}=t_{0}$, using the initial material points $\mathbf{M}(t_{0})$ and apparent motion cues $\mathbf{A}(t_{0}, t_{1}) = \left\{\mathbf{S}(t_{0}), \mathbf{S}(t_{1})\right\}$ to predict material points $\hat{\mathbf{M}}(t_{1})$; then $\hat{\mathbf{M}}(t_{1})$ and $\mathbf{A}(t_{1}, t_{2})= \left\{\mathbf{S}(t_{1}), \mathbf{S}(t_{2})\right\}$ will be used to predict $\hat{\mathbf{M}}(t_{2})$, and so forth. In such a sequential prediction way, we will recover the 3D material point trajectory $\left\{\mathbf{M}(t_{q}) \right\}_{q=0}^{T-1}$ over a cardiac cycle of $T$ phases. Since we always deform an identical set of material points, \ie, $\mathbf{M}(t_{0})$, the correspondence of each material point between different time points is established automatically.

\section{Experiments}
\subsection{Dataset and Pre-Processing}
A large public 3D CMR dataset~\cite{bai2015bi} of $1,331$ normal subjects was used in our experiments for 3D left ventricle heart wall geometry and motion simulation, and evaluation of 3D motion recovering accuracy. Each subject in the dataset has high resolution (HR) segmentation masks at ED and ES phases.
The HR data was scanned by 3D high spatial resolution of $1.25\times1.25\times2\; mm^{3}$ CMR protocols~\cite{de2014population}.

To simulate heart wall geometry and motion, the inputted sparse point cloud in Step $\boldsymbol{\mathit{1}}$ was generated and pre-processed in the same way as in~\cite{ye2023neural}. 
Each data set at ED phase was rotated around the $z$-axis to align the center line of LV and RV with the $y$-axis, such that the points lie in the same coordinate system as we defined in Sec.~\ref{sec31preliminary}. The same rotation angle was applied to the data at ES phase.
We set $N_{u}=N_{v}=50$, $N_{w}=9$ ($w = 0, \dots, 8$), $T=20$. 
The odd wall layers ($N_{w, o}=4$) were used to generate the SPAMM datapoints and even wall layers ($N_{w, e}=5$) were used as material points.

For generation of multi-planar SPAMM datapoints, we followed standard clinical 2D tagged MRI scanning approaches. We sampled SPAMM datapoints at 3 LAX and 10 SAX planes. For each subject, these planes were \texttt{fixed} over time and were ensured to span the spatial extent of the whole LV at ED.
Because of the heart wall 3D motion, some SPAMM datapoints on a certain \texttt{fixed} imaging plane could disappear or reappear at subsequent time points.
At each temporal point $t_{q}$, we only used the points $\mathbf{S}(t_{q+1})$ which have a corresponding point in $\mathbf{S}(t_{q})$~\cite{park1996analysis}. The varying numbers of \textit{active} SPAMM datapoints were fixed as $N_{s}=3200$ by using farthest point sampling (FPS)~\cite{qi2017pointnet++}. 
For pre-processing of 3D motion recovering, we centered each material point sets $\mathbf{M}(t_{0})$ at $(0, 0, 0)$ by subtracting their center coordinates $\boldsymbol{c}(t_{0})$, and linearly normalized $x$, $y$, $z$ coordinates to $[-1.5, 1.5]$ with the scales $\boldsymbol{a}(t_{0})$. The SPAMM datapoints $\mathbf{S}(t_{q})$ and ground truth material points $\mathbf{M}(t_{q})$ were first translated by subtracting $\boldsymbol{c}(t_{0})$ and then linearly normalized using the same $\boldsymbol{a}(t_{0})$. In this way, all the material points and SPAMM datapoints were normalized in the same coordinate system. 

Since we were dealing with a large dataset of time sequences ($T=20$), we randomly selected 500 subjects from the 3D CMR dataset~\cite{bai2015bi}.  These 500 subjects were then randomly split into 200, 100 and 200 subjects as the training, validation and test sets, respectively.

\subsection{Evaluation Metrics}
To quantitatively evaluate the 3D motion recovery accuracy, we use the mean absolute error (MAE) of Euclidean distance between predicted material points $\hat{\mathbf{M}}(t_{q})$ and ground truth $\mathbf{M}(t_{q})$.
In addition, we evaluated the self-intersection (SI) ratio~\cite{sun2022topology, ye2023neural} for each Lagrangian motion field~\cite{ye2023sequencemorph} deformed heart wall layer mesh.
If the estimated Lagrangian motion is smooth with inverse smooth,  there should be no manifold folding in the deformed mesh.

\subsection{Baseline Methods}
We compared our method with a state-of-the-art meshless deformable model (MDM)~\cite{wang2008meshless, wang2015meshless}, which is a conventional iterative optimization-based method. We reimplemented MDM with C++.
To the best of our knowledge, we are the first to develop a learning-based approach to recover 3D true heart wall motion from 2D apparent motion cues. 
Therefore, we lack learning-based baseline methods for comparison in this task.
We also compared our method with NDM~\cite{ye2023neural}, to show the failure of NDM on material point correspondence recovery.

\subsection{Implementation Details}
We implemented $\upsilon$NDMs based on Point Transformer~\cite{zhao2021point} with PyTorch. Any batch normalization~\cite{ioffe2015batch} used in Point Transformer was removed. In the point cross-attention layer, we set $k=64$ for nearest neighbouring key points assigning.
In the loss function, we set the hyper-parameters $\lambda_{d}=0.1$, $\lambda_{s}=0.05$ via grid search.
Adam optimizer was used with a learning rate of $1e^{-4}$ to train our network $\Psi$. All models were trained on an NVIDIA A100 GPU with $E_{1}=1000$,  $E_{2}=300$ epochs. 

\subsection{Results}
\subsubsection{3D Motion Recovery Accuracy}

\begin{table}
\begin{center}
\resizebox{0.90\columnwidth}{!}{
\begin{tabular}{l|c|c|c}
\hline
Method &MAE ($mm$) $\downarrow$ & SI $\downarrow$ & Time ($s$) $\downarrow$\\
\hline\hline
MDM &$1.180\pm0.314$ &$0.139\pm0.086$ &$5.685\pm0.120$ \\
NDM &$1.532\pm0.243$ &$0.187\pm0.123$ &$\mathbf{0.110}\pm0.035$\\
Ours &$\mathbf{0.954}\pm0.262$ &$\mathbf{0.096}\pm0.103$ &$0.147\pm0.045$\\
\hline
\end{tabular}}
\end{center}
\caption{Performance comparison for 3D motion recovery. We compared the mean absolute error (MAE) in a cardiac cycle, ratio of self-intersection (SI) faces on the Lagrangian motion field deformed heart wall mesh and running time.}
\label{table1}
\end{table}

\begin{figure}
\begin{center}
\includegraphics[width=0.6\linewidth]{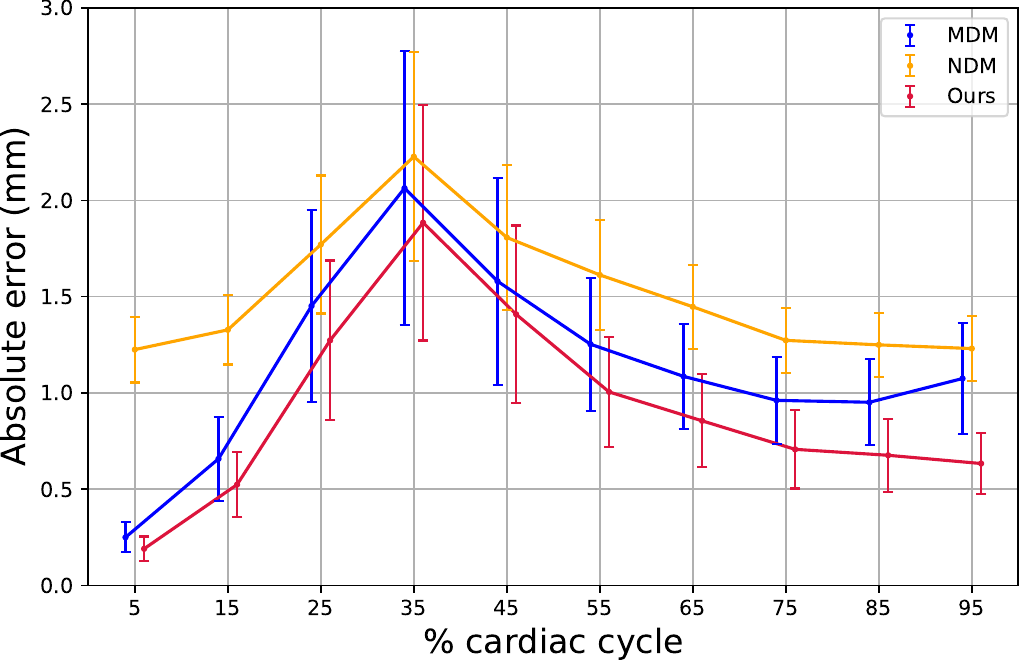}
\end{center}
   \caption{Mean and standard deviation of the absolute errors over a cardiac cycle for different 3D motion recovery methods.
   }
\label{fig7}
\end{figure}

\begin{figure*}
\begin{center}
\includegraphics[width=1.0\linewidth]{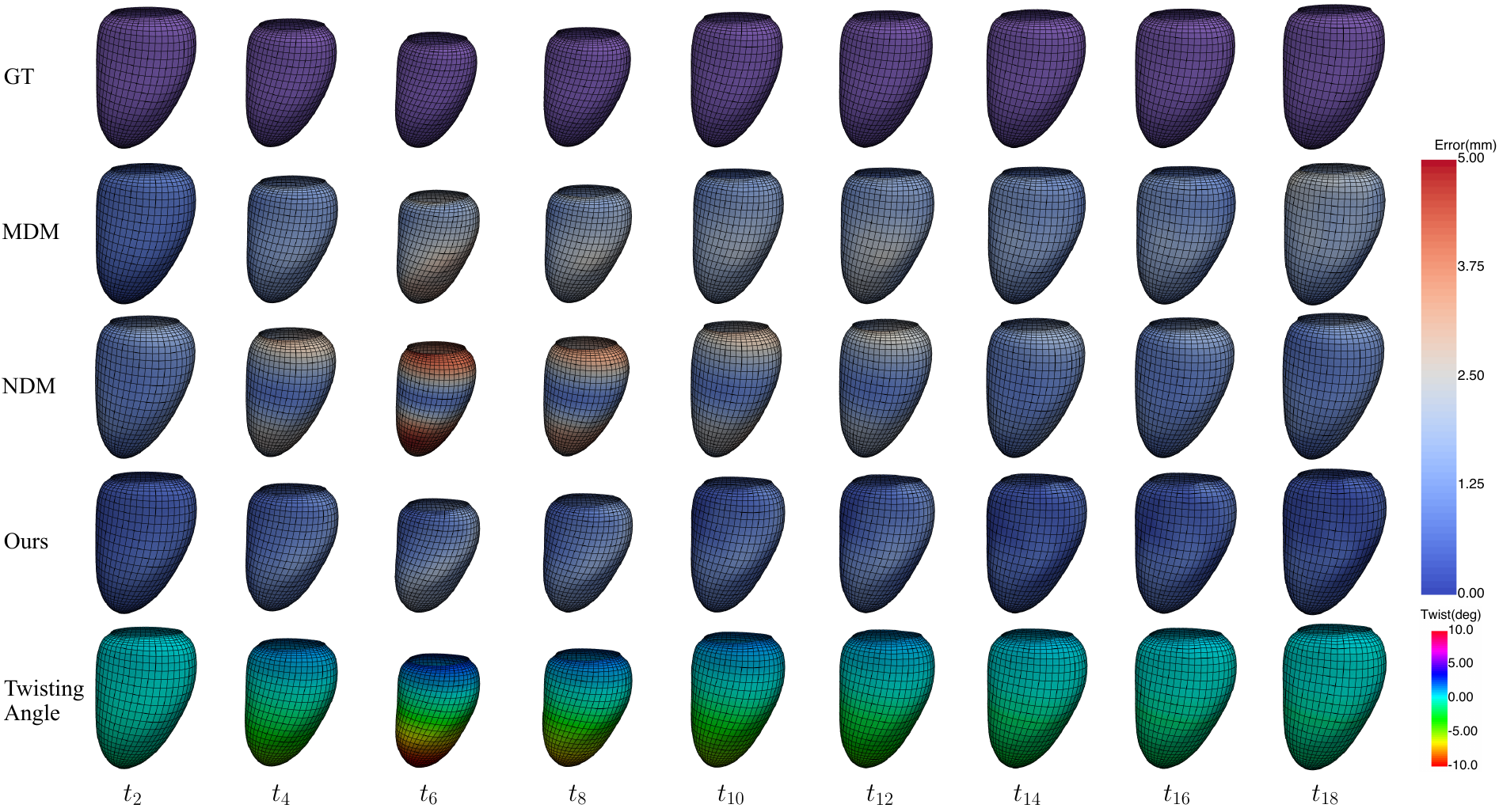}
\end{center}
   \caption{Results of 3D heart wall motion recovery over a cardiac cycle ($T=20$). For each method, we show the Lagrangian motion deformed mesh of the middle wall surface, with colors indicating the Euclidean distance from the prediction to ground truth (GT). The last row shows regional twisting angles over different cardiac phases recovered by our method.
   }
\label{fig8}
\end{figure*}

In Table~\ref{table1}, we show quantitative results of 3D motion recovery from 2D apparent motion cues. The MAE was averaged over a whole cardiac cycle.  Mean and standard deviation of the absolute errors over a cardiac cycle are shown in Fig.~\ref{fig7}. We also show an example in Fig.~\ref{fig8}.
MDM is an iterative method, whose performance relies on tuning of hyper-parameters, \eg, the phyxel kernel radius. Hyper-parameters tuned on the training dataset cannot guarantee a global optimum solution for each test data set.
NDM only uses the inner and outer wall surface points to reconstruct the heart wall geometry. The temporal correspondence provided by the SPAMM datapoints is ignored. That is the reason why NDM fails to accurately recover the dense material point temporal correspondence.
Our method uses a hybrid point transformer to fuse 2D apparent motion cues into 3D true motion hints and map the hints to material point true motion, which can be captured by the deformation parameters of $\upsilon$NDM. The efficient usage of temporal SPAMM datapoint correspondence results in accurate 3D true motion recovery.
The Lagrangian motion field deformed wall mesh is nearly smooth, with self-intersection face ratio close to 0.

\begin{table}
\begin{center}
\resizebox{0.62\columnwidth}{!}{
\begin{tabular}{l|c|c}
\hline
$k$ &MAE ($mm$) $\downarrow$ & SI $\downarrow$  \\
\hline\hline
32 &$1.246\pm0.265$ &$\mathbf{0.075}\pm0.053$\\
64 &$\mathbf{0.954}\pm0.262$ &$0.096\pm0.103$\\
96 &$1.067\pm0.324$ &$0.095\pm0.070$\\
\hline
\end{tabular}}
\end{center}
\caption{Ablation of the number of $k$-NN key points assigned in the point cross-attention layer.}
\label{table2}
\end{table}

\subsubsection{Running Time Analysis}
In Table~\ref{table1}, we report the average inference time for 3D motion recovery at each time point in the cardiac cycle. 
While MDM uses an Intel Xeon CPU, the learning-based methods utilize both CPU and GPU during inference.
We note that our learning-based method, $\upsilon$NDM, is much faster than the conventional iteration-based method, MDM. 

\begin{table}
\begin{center}
\resizebox{1.0\columnwidth}{!}{
\begin{tabular}{l|c|c|c|c|c|c}
    \hline
    Model &$\boldsymbol{q}_{g}$ &$\boldsymbol{q}_{d}$ & S2 & Mixed &MAE ($mm$) $\downarrow$&SI $\downarrow$\\
    \hline\hline
    A1 &\xmark & \cmark &\cmark & \cmark &$1.079\pm0.320$ &$0.101\pm0.080$\\
    A2 &\cmark & \xmark &\cmark & \cmark &$1.216\pm0.299$ &$\mathbf{0.060}\pm0.049$\\
    A3 &\cmark & \cmark  &\xmark & \cmark &$1.310\pm0.272$ &$0.074\pm0.055$\\
    A4 &\cmark & \cmark  &\cmark & \xmark &$1.077\pm0.290$ &$0.117\pm0.083$\\
    Ours &\cmark & \cmark &\cmark & \cmark &$\mathbf{0.954}\pm0.262$ &$0.096\pm0.103$\\
    \hline
    \end{tabular}}
    \end{center}
    \caption{Ablation of global and local deformations in $\upsilon$NDMs (A1 and A2), training Stage \RNum{2} (S2) for accumulation error reducing (A3), and separated or mixed usage of $\mathbf{S}(t_{q+1})$ on SAX and LAX views in the point cross-attention layer (A4).}
    \label{table3}
\end{table}

\subsubsection{Ablation Study}
We conducted ablation studies to investigate the effects of the main components in our approach.

From Table~\ref{table2}, when the $k$-NN number is too small, less neighboring apparent motion cues are fused, which harms the true motion hints generation. When $k$ is too large, more irrelevant neighboring apparent motion cues can mislead the generation of true motion hints.

From Table~\ref{table3}, neither using local (A1) nor global (A2) deformation parameters in $\upsilon$NDMs is sufficient for recovering complex 3D heart wall motion;
training Stage \RNum{2} (A3) helps to reduce accumulation error in sequential recovery of 3D true motion over time;
the mixed usage of $\mathbf{S}(t_{q+1})$ performs better than separated usage (A4). 
The `separated' usage means that
for SAX views, we take the $x$ and $y$ coordinates of $\mathbf{S}(t_{q+1})$; for LAX views, we only take the $z$ coordinate of $\mathbf{S}(t_{q+1})$. 
While in the conventional deformable model-based approach~\cite{park1996analysis}, this is a common operation for multi-planar apparent motion cue fusion, using point cross-attention and MLP, we can make better use of the cues provided by SPAMM datapoints in both SAX and LAX views.

\section{Conclusion}
Our $\upsilon$NDMs can learn to recover 3D true motion from 2D apparent motion cues for high-dimensional regional heart wall motion sequences.  
Using $\upsilon$NDMs, we can simulate complex heart wall geometry and motion, including twisting, over a full cardiac cycle. 
The design of a hybrid point transformer was demonstrated to be effective for solving specific challenges in our task. 
We expect more applications of our approach to other similar 3D motion recovery tasks from 2D apparent motion cues.

{
    \small
    \bibliographystyle{ieeenat_fullname}
    \bibliography{main}
}
\clearpage
\setcounter{page}{1}
\maketitlesupplementary

\section{Definition of MAE}
The quantitative metric for 3D motion recovery accuracy evaluation we used is the mean absolute error (MAE). It is the average of Euclidean distance ($\ell_{2}$ norm) between predicted material points $\hat{\mathbf{M}}(t_{q})$ and ground truth $\mathbf{M}(t_{q})$ over a full cardiac cycle excluding the starting time point:
\begin{equation}
    \text{MAE}=  \frac{\sum_{q=1}^{T-1}\left\|\hat{\mathbf{M}}(t_{q})- \mathbf{M}(t_{q})\right\|_{2}}{T-1},
\end{equation}
where $T$ is the total number of phases over a cardiac cycle.

\section{Temporal Interpolation Scalars}
The temporal interpolation scalars $s(t_{q})$ used in Eq.~(\ref{eqmtq}) in the main text are set as follows: 
\begin{equation}
\begin{aligned}    
    s_{x}(t_{q})= [&\mathbf{0.000}, 0.090, 0.150, 0.350, 0.550, 0.750, \mathbf{1.000}, 
    \\&0.920, 0.780, 0.650, 0.580, 0.540, 0.410, 0.370, 
    \\&0.240, 0.210, 0.180, 0.160, 0.120, 0.080],
\end{aligned}
\end{equation}

\begin{equation}
\begin{aligned}    
    s_{y}(t_{q})= [&\mathbf{0.000}, 0.080, 0.180, 0.380, 0.580, 0.780, \mathbf{1.000}, 
    \\&0.980, 0.740, 0.620, 0.550, 0.520, 0.480, 0.350, 
    \\&0.220, 0.190, 0.180, 0.160, 0.110, 0.080],
\end{aligned}
\end{equation}

\begin{equation}
\begin{aligned}    
    s_{z}(t_{q})= [&\mathbf{0.000}, 0.100, 0.200, 0.400, 0.600, 0.800, \mathbf{1.000}, 
    \\&0.920, 0.880, 0.650, 0.380, 0.335, 0.325, 0.315, 
    \\&0.3000, 0.290, 0.280, 0.220, 0.090, 0.070].
\end{aligned}
\end{equation}

Note that, except for ED ($t_{ed} = t_{0}$) and ES ($t_{es}= t_{6}$), $s(t_{q})$ are different for $x$, $y$, and $z$ at each time point. We do this to introduce nonlinear deformations into the generated meshes $\left\{\mathbf{M}(t_{q}) \right\}_{q=0}^{19}$ across time.

\section{Quality Control of Synthetic Data}
\begin{figure*}
\begin{center}
\includegraphics[width=1.0\linewidth]{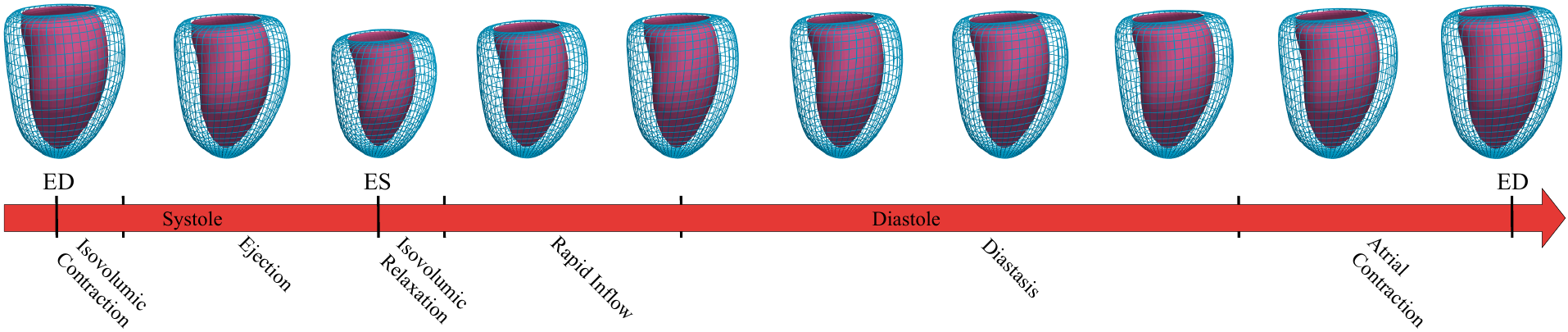}
\end{center}
   \caption{Simulated heart wall shapes of the left ventricle (LV) across a full cardiac cycle. For each 3D wall shape, blue is the epicardial surface; red is the endocardial surface. Along the time axis, we show different stages of a full cardiac cycle. ED: end diastole. ES: end systole.
   }
\label{figs1}
\end{figure*}

\begin{figure*}
\begin{center}
\includegraphics[width=1.0\linewidth]{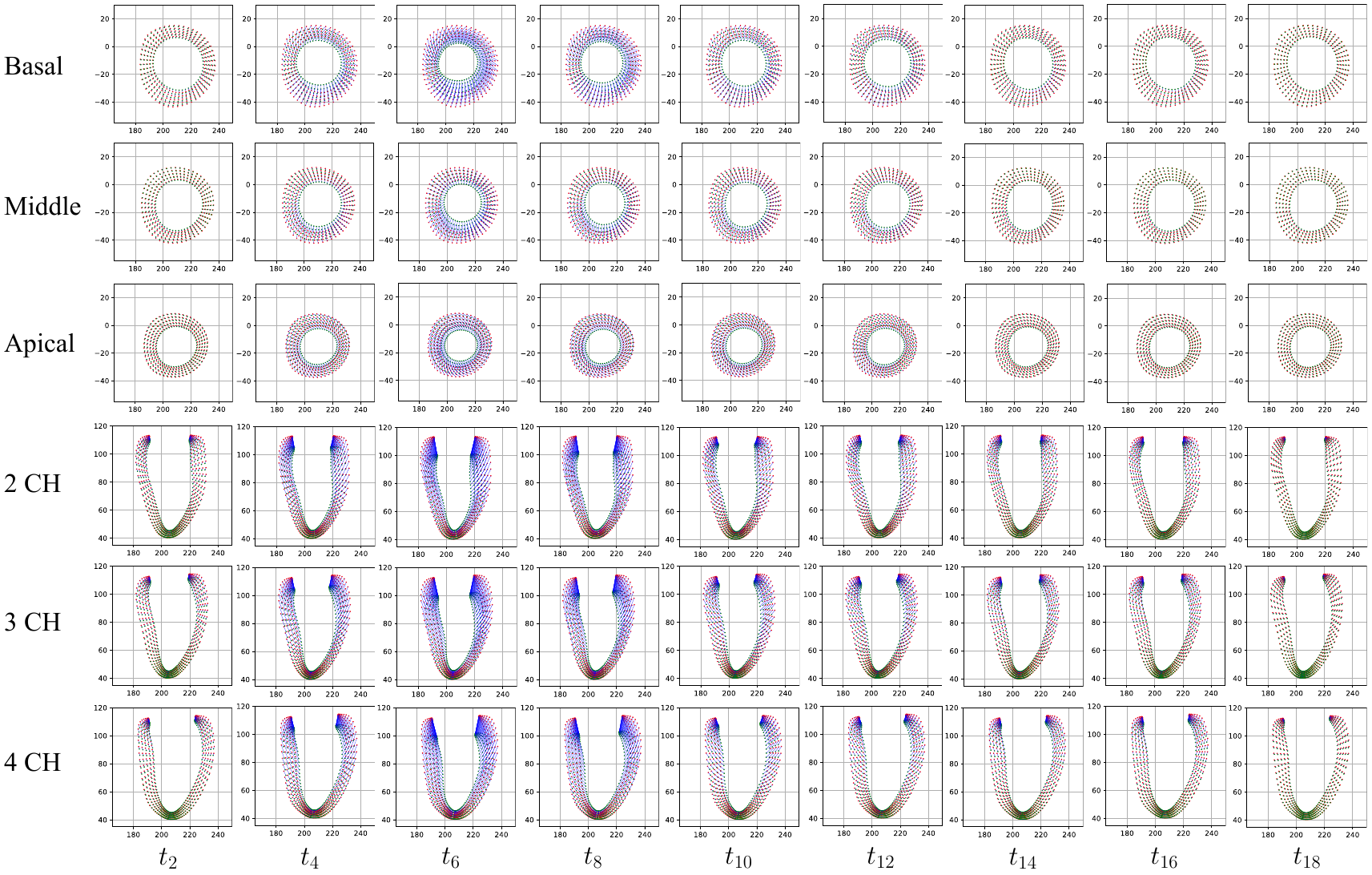}
\end{center}
   \caption{Apparent Lagrangian motion of computed SPAMM datapoints on the short axis (SAX) and long axis (LAX) views. For SAX views, we show the imaging planes in basal, middle and apical regions when looking down from base to apex. For LAX views, we show the imaging planes in 2 chamber (CH), 3 CH and 4 CH directions. Red are reference points at $t_{0}$; green are moving points; blue lines show the correspondences.
   }
\label{figs2}
\end{figure*}

One of our contributions in this work is the 3D heart wall geometry and motion simulation framework. 
While we use a 3D CMR dataset~\cite{bai2015bi} of normal subjects and data interpolation defined in Eq.~(\ref{eqmtq}) in the main text to achieve this goal, the missing heart wall geometry shapes at certain cardiac phases in this dataset can make the synthetic data deviate from real data distribution. 
To ensure the synthetic heart walls deform in a realistic way, we asked a radiologist (\textbf{L.A.}) with more than 40 years' experience on cardiac imaging to do the quality control. 
In Fig.~\ref{figs1}, we show examples of the simulated left ventricle (LV) heart wall shapes which deform across a full cardiac cycle. 
In Fig.~\ref{figs2}, we show corresponding apparent Lagrangian motion of computed SPAMM datapoints on the short axis (SAX) and long axis (LAX) views.
In the supplementary folder \verb'Simulation_Example', we show these examples in dynamic videos. 
Overall, the deformation of synthetic heart walls should follow these rules:
\begin{enumerate}[label=(\arabic*)]
\item The apex of the heart wall is nearly immobile.
\item The differential torsional (or twisting) motion between the different SAX levels should be seen, along with its temporal behavior, with most rotation seen relatively early in systole and then (``back'') mostly in the quite early part of diastole. The regional rotations should not completely uniform around the heart; \ie, it does not move like rigid rings.
\item The three phases of diastole (rapid inflow, diastasis and atrial contraction) near the base of the heart on the LAX view images should be clearly seen. The atrial contraction pulls up on the base of the LV, which should be seen on the LAX view images, with a nonuniform distribution (motion mostly at the LV base).
\item There should be a gradient of the longitudinal displacement from base to apex, as seen in the corresponding LAX view blue line segment lengths.
\item The apparent motion on different LAX views should look very similar qualitatively.
\end{enumerate}
The synthesized instances which break any of the above rules were corrected or removed.

\section{More Detailed Results}
In Fig.~\ref{figs3}, we show more detailed 3D heart wall motion recovery results. 
In the supplementary folder \verb'3D_Regional_Wall_Motion_Recovery_Example', we show these examples in dynamic videos. 
From these results, we can see that our method can accurately recover the volumetric regional heart wall motion from 2D apparent motion cues provided by MR SPAMM-tagging images. 

\begin{figure*}
\begin{center}
\includegraphics[width=1.0\linewidth]{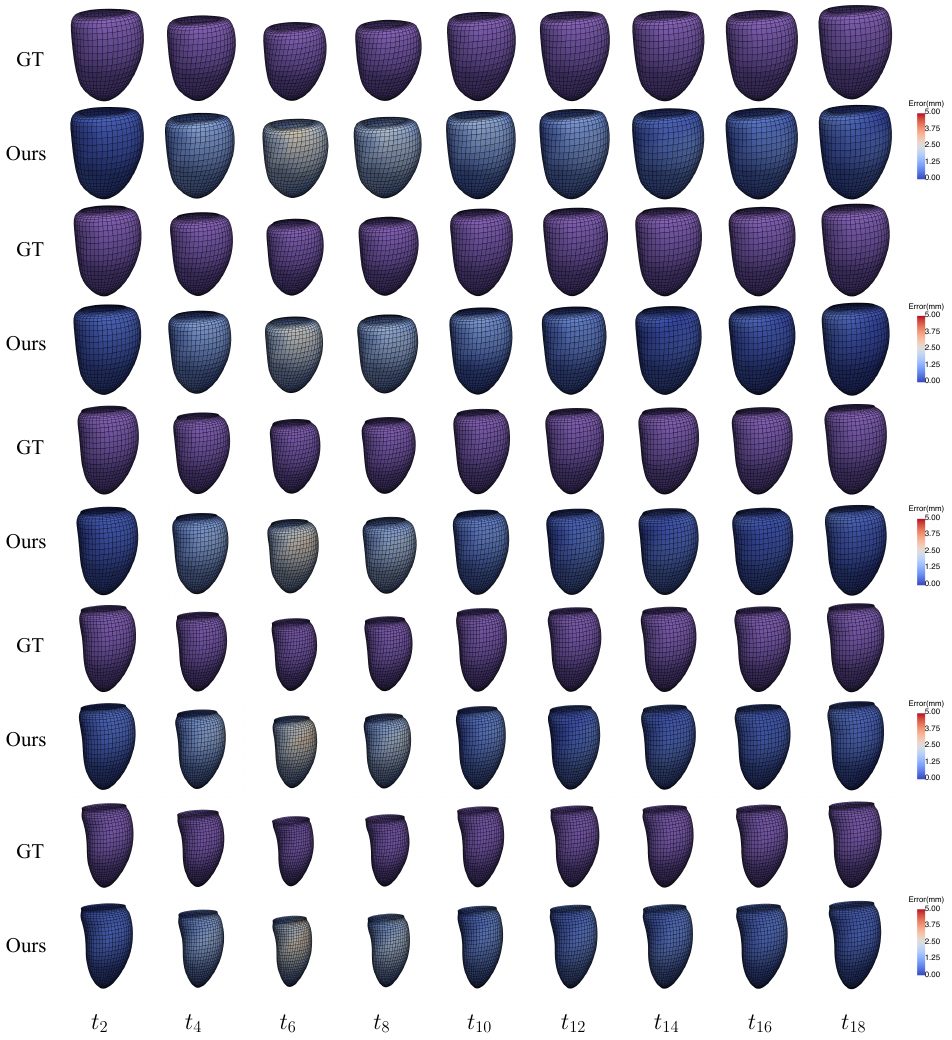}
\end{center}
   \caption{More results of 3D heart wall motion recovery over a cardiac cycle ($T=20$). From top to bottom, we show results of 5 heart wall layers, \ie, from epicardial surface to endocardial surface. 
   }
\label{figs3}
\end{figure*}


\end{document}